\documentclass[12pt]{article}
\usepackage{latexsym}

\def\bs{\begin{subequations}}
\def\es{\end{subequations}}

\catcode`\@=11

\newtoks\@stequation
\def\subequations{\refstepcounter{equation}
  \edef\@savedequation{\the\c@equation}%
  \@stequation=\expandafter{\theequation}
  \edef\@savedtheequation{\the\@stequation}
  \edef\oldtheequation{\theequation}%
  \setcounter{equation}{0}%
  \def\theequation{\oldtheequation\alph{equation}}}

\def\endsubequations{\setcounter{equation}{\@savedequation}%
  \@stequation=\expandafter{\@savedtheequation}%
  \edef\theequation{\the\@stequation}\global\@ignoretrue}

\makeatletter
        \renewcommand{\theequation}{\thesection.\arabic{equation}}%
        \@addtoreset{equation}{section}%
\makeatother

\renewcommand{\thefootnote}{\fnsymbol{footnote}}

\begin{document}

\begin{titlepage}

 April 14, 2011

\begin{center}        \hfill   \\
            \hfill     \\
                                \hfill   \\

\vskip .25in

{\large \bf Tachyons and Gravitational Cherenkov Radiation \\}

\vskip 0.3in

Charles Schwartz\footnote{E-mail: schwartz@physics.berkeley.edu}

\vskip 0.15in

{\em Department of Physics,
     University of California\\
     Berkeley, California 94720}
        
\end{center}

\vskip .3in

\vfill

\begin{abstract}
We calculate the rate at which a free tachyon (faster than light particle) 
would emit gravitational radiation.  It is very small.

\end{abstract}

\vfill

\end{titlepage}

\renewcommand{\thefootnote}{\arabic{footnote}}
\setcounter{footnote}{0}
\renewcommand{\thepage}{\arabic{page}}
\setcounter{page}{1}

In a recent work \cite{CS} I have suggested considering that faster-than-light 
particles (tachyons) might exist in the universe and could be a significant 
element in cosmological models. 

Travelling faster than c, the speed of light, they would then be 
expected to emit gravitational radiation - analogous to the 
electromagnetic radiation (Cherenkov radiation) observed when 
electrically charged particles travel at high speeds through a medium 
in which light waves propagate at speeds less than c. 

We start this calculation with formulas given by Weinberg \cite{SW1}  for 
gravitational radiation from a given source.

\begin{equation}
\frac{dE}{d\Omega} = 2 G \int d \omega\;\omega^{2}[T^{\lambda \nu 
*}(\textbf{k},\omega) T_{\lambda \nu}(\textbf{k},\omega) - 
\frac{1}{2} |T^{\lambda}_{\lambda}(\textbf{k},\omega)|^{2}],\label{a1}
\end{equation}
for the total energy emitted per solid angle in the direction 
$\hat{\textbf{k}}$, with the Fourier transformed energy momentum 
tensor,
\begin{equation}
T_{\mu \nu} (\textbf{k},\omega) \equiv \frac{1}{2\pi}\int 
d^{4}x\;e^{i k\cdot x}\;T_{\mu \nu}(\textbf{x},t).\label{a2}
\end{equation}

For a free tachyon we have ,
\begin{eqnarray}
T_{\mu \nu}(x) = m \int d \tau\; \dot{\xi}_{\mu}\;\dot{\xi}_{\nu} \;
\delta^{4}(x-\xi(\tau))\label{a3} \\
\xi^{\mu}(\tau) = (\gamma \tau, \gamma \textbf{v}\;\tau), \;\;\;\;\; 
\gamma = 1/ \sqrt{v^{2}-1}, \;\;\;\;\;\dot{\xi}^{\mu}\dot{\xi}_{\mu} 
= -1.\label{a4}
\end{eqnarray}
This gives us the Fourier transform,
\begin{equation}
T_{\mu \nu} (\textbf{k},\omega) = m\;\dot{\xi}_{\mu}\dot{\xi}_{\nu} \;
\delta(\gamma \omega (1-\hat{\textbf{k}}\cdot \textbf{v})).\label{a5}
\end{equation}
The argument of this  delta-function is the Lorentz invariant 
quantity $\dot{\xi}_{\mu}\;k^{\mu}$. It tells us that the radiation 
goes out along a cone, just as with familiar Cherenkov radiation; and 
this can only happen for velocities $v$ that are greater than $c=1$.

Now we put (\ref{a5}) into (\ref{a1}) and get,
\begin{equation}
\frac{dE}{d\Omega} = 2G \int d\omega \;\omega^{2}\; \frac{m^{2}}{2} 
\delta (\gamma \omega (1-\hat{\textbf{k}}\cdot 
\textbf{v}))\;\delta(0).\label{a6}
\end{equation}
We make sense out of this by noting that
\begin{equation}
\delta(0) = \frac{1}{2\pi} \int d \tau e^{i (0)\tau} = \frac{\Delta 
\tau}{2\pi} = \frac{\Delta t}{2 \pi \gamma}.\label{a7}
\end{equation}
where $\Delta t$ is the time interval over which we observe this 
process of radiation.

Now we integrate over all angles and get the rate of energy emission,
\begin{equation}
\frac{\Delta E}{\Delta t} = G\;\frac{m^{2}}{\gamma^{2}\;v} \int 
d\omega\;\omega.\label{a8}
\end{equation}

We need to introduce some cut-off for the integral over $\omega$; and 
this we take from the basic quantum relation, $E = \hbar \omega$, 
where this E of the emitted gravitational quantum cannot be more than 
the total energy E of the tachyon.  Thus we end with,
\begin{equation}
\frac{\Delta E}{\Delta t} \sim \frac{G}{2}\;\frac{m^{2}}{\gamma^{2}\;v} 
(E/\hbar)^{2}.
\end{equation}
Noting that $E = m\gamma$, and putting in the factors of c, we get 
our final result,
\begin{equation}
\frac{\Delta E}{\Delta t} \sim 
\frac{G}{2}\;\frac{m^{4}c^{4}}{\hbar^{2}\;v}.\label{a9}
\end{equation}

Putting in numbers we have:
\begin{equation}
\frac{\Delta E}{\Delta t} \sim (\frac{mc^{2}}{eV})^{4}\;(\frac{c}{v})\; 10^{-41}\;
eV/sec.\label{a10}
\end{equation}
Suppose we guess that the neutrino is a tachyon, with mass somewhat 
less than one electron Volt. Over the course of cosmic time, $\sim 10^{17} 
sec$, the gravitational radiation it will have emitted will be 
incredibly small at around $10^{-24}$ eV.

I thank K. Bardakci for some helpful discussions on this topic.

\end{document}